\documentclass[10pt]{iopart}
\usepackage{latexsym}
\usepackage{amsfonts}
\usepackage{amsbsy}
\usepackage{amssymb}   
\usepackage{mathrsfs}
\usepackage[mathscr]{euscript}
\usepackage{accents}
\usepackage{hyperref}
\hypersetup{colorlinks=true,linkcolor=blue,urlcolor=blue,citecolor=blue}

\newcommand{\eqref}[1]{(\ref{#1})}
\usepackage{subeqn}

\usepackage[makeroom]{cancel}

\usepackage{tensor}
\usepackage{color}
\usepackage{cite}

\usepackage{lipsum}
\usepackage{relsize}
\usepackage{color}
\usepackage{nameref}
\usepackage{tikz}

\usepackage{tikz-cd}
\usepackage{xcolor}

\begin{document}

\title[Classical and quantum trace-free Einstein cosmology]
{Classical and quantum trace-free Einstein cosmology}

\author{Merced Montesinos$^{1}$\footnote{Author to whom any correspondence should be addressed.}\,\href{https://orcid.org/0000-0002-4936-9170}{\includegraphics[scale=0.05]{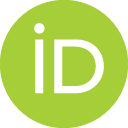}}, Abdel P\'erez-Lorenzana$^{1}$\,\href{https://orcid.org/0000-0001-9442-3538}{\includegraphics[scale=0.05]{ORCIDiD_icon128x128.png}},\\ Jorge Meza$^{1}$\,\href{https://orcid.org/0009-0002-5022-219X}{\includegraphics[scale=0.05]{ORCIDiD_icon128x128.png}},  and Diego Gonzalez$^{1,2}$\,\href{https://orcid.org/0000-0002-0206-7378}{\includegraphics[scale=0.05]{ORCIDiD_icon128x128.png}}}

\address{$^1$ Departamento de F\'{\i}sica, Cinvestav, Avenida Instituto Polit\'ecnico Nacional 2508, 	San Pedro Zacatenco, 07360 Gustavo A. Madero, Ciudad de M\'exico, M\'exico.}
\address{$^2$ Escuela Superior de Ingenier\'ia Mec\'anica y El\'ectrica, Instituto Polit\'ecnico Nacional, Unidad Profesional Adolfo L\'opez Mateos, Zacatenco,  07738 Gustavo A. Madero, Ciudad de M\'exico, M\'exico}

\eads{\mailto{merced.montesinos@cinvestav.mx}, \mailto{abdel.perez@cinvestav.mx}, \mailto{jorge.meza@cinvestav.mx}, and \mailto{dgonzalezv@ipn.mx}}


\begin{abstract}

Trace-free Einstein gravity, in the absence of matter fields and using the Friedmann-Robertson-Walker (FRW) metric, is solvable both classically and quantum mechanically. This is achieved by using the conformal time as the time variable and the negative or positive of the inverse of the scale factor as configuration variable to write the classical equation of motion, which turns out to be the one of a free particle ($k=0$), a harmonic oscillator ($k=1$), and a repulsive oscillator ($k=-1$) in the real half-line. In all cases, the observable identified as the cosmological constant is six times the Hamiltonian. In particular, for a closed Universe ($k=1$), spacetime exhibits a cyclic evolution along which the scalar curvature is constant and finite, thereby avoiding singularities. The quantum theory is reached by using canonical quantization. We calculate the spectrum of the observable corresponding to the cosmological constant. Remarkably, for the closed Universe ($k=1$), the spectrum is discrete and positive while for flat ($k=0$) and open ($k=-1$) universes, the spectra are continuous. Heisenberg's uncertainty principle imposes limitations on the simultaneous measurement of the Hubble expansion (momentum variable) and the configuration variable. We also report the observable identified as the cosmological constant for inflaton, phantom and perfect fluids coupled to trace-free Einstein gravity in the FRW metric.

\end{abstract}



\section{Introduction}

Nearly a century ago, Einstein proposed an alternative theory of gravity~\cite{Einstein_1919,Einstein_1952} in which, instead of using the full equations of general relativity, only their trace-free part is considered. In four dimensions, the trace-free Einstein equations are~\cite{Einstein_1927}
\begin{eqnarray}\label{TFEM}
R_{\mu\nu} - \frac14 R g_{\mu\nu} = \kappa^2 \Big( T_{\mu\nu} - \frac14 T g_{\mu\nu}\Big),
\end{eqnarray}
where $R_{\mu\nu}=R^{\theta}{}_{\mu\theta\nu}$ and $R=g^{\mu\nu} R_{\mu\nu}$ are the Ricci tensor and the scalar curvature, respectively, both derived from the components of the Riemann tensor $R^{\theta}{}_{\mu\sigma\nu}$ calculated  with the Levi-Civita connection compatible with the metric $g_{\mu\nu}$, and  $T_{\mu\nu}$ is the energy-momentum tensor of matter fields. A crucial difference between trace-free Einstein gravity and general relativity is that the cosmological constant does not appear in the equations of motion of the former (see equation~\eqref{TFEM}), whereas in the latter, it is present as a parameter in its equations of motion. However, due to second Bianchi identity and the assumption of the conservation of the energy-momentum, a constant of motion naturally emerges in trace-free Einstein gravity, which is identified as the cosmological constant. This fact is often underestimated or overlooked; however, it has profound implications for our understanding of gravitational phenomena~\cite{Ellis_2011,Ellis_2014}. Despite this, trace-free Einstein gravity has only recently begun to receive attention. In particular, the works~\cite{MontGonz_2023,Montesinos_2025} introduce action principles for this theory that do not involve any unimodular condition to produce the equations~\eqref{TFEM}. Specifically, such action principles express trace-free Einstein gravity as either a real BF-type theory \cite{MontGonz_2023} or as a complex BF-type theory \cite{Montesinos_2025}. These formulations are particularly valuable, especially given their potential usefulness in the quantization program of the gravitational field. Because of the absence of the unimodular condition, trace-free Einstein gravity is not unimodular gravity. In particular, there is no action principle in the metric formalism for trace-free Einstein gravity, as happens for unimodular graviy~\cite{Anderson_1971,Weinberg_1989}. The best we have is a metric formulation that in addition involves an auxiliary metric, resulting in a constrained bigravity theory~\cite{MontGonz_2024}. One of the key features of this formulation is that energy-momentum conservation is not assumed a priori; instead, it arises naturally from the equations of motion.

Motivated by the lack of an action principle for trace-free Einstein gravity in the metric formalism, we investigate the equations of motion of the theory coupled to matter fields (inflaton $\Phi$, phantom $\varphi$, and perfect fluids) in the Friedmann-Robertson-Walker (FRW) metric to get insights into the full theory. In the FRW metric, we find that the trace-free Einstein equations reduce to just one equation that exhibits no residual gauge symmetry, in contrast to general relativity in the same metric. Additionally, we show how the constant of motion (identified as the cosmological constant) emerges, and we provide several Lagrangian and Hamiltonian formulations of the equation of motion when there are no matter fields. Furthermore, using the conformal time as the evolution parameter, the equation of motion acquires the simple forms of a free particle ($k=0$), a harmonic oscillator ($k=1$), and a repulsive oscillator ($k=-1$) in the real half-line, and in the absence of matter fields. Therefore, in this case, the problem is solvable both classical and quantum mechanically. In the case of a closed Universe ($k=1$), spacetime undergoes a cyclic classical evolution in the conformal time. Throughout this cyclic evolution, both the constant of motion and the scalar curvature remain finite and positive, and thus there is no singularity. 

The quantum theory in the absence of matter fields is obtained by making the canonical quantization of the Hamiltonian formulation that employs the conformal time as the evolution parameter. Since there are no constraints classically, the evolution of quantum states is governed by the usual Schrödinger equation rather than by a quantum constraint as happens for the FRW mini-superspace of general relativity~\cite{DeWitt_1967,Misner_1969} (see also~\cite{Tipler,Moniz}). We calculate the spectrum of the observable corresponding to the cosmological constant for the three cases of the spatial topology. Remarkably, for the closed spacetime ($k=1$), this spectrum is discrete and positive. For $k=0$ and $k=-1$, the spectra are continuous. Furthermore, due to the fact that the momentum variable is the expansion rate, Heisenberg's uncertainty relation imposes intrinsic limitations on the simultaneous measurements of the expansion rate and the scale factor.   

\section{Trace-free Einstein cosmology in the cosmic time}

We consider the FRW metric $g$ expressed in local coordinates $x^{\mu}=(t,x^a)= (t,r,\theta,\phi)$ as
\begin{eqnarray}\label{FRWspacetime}
g = - dt^2 +  a^2(t) \left [ \frac{d r^2}{ 1-k r^2 }+ r^2 \left ( d \theta^2 + \sin^2{\theta} d \phi^2 \right )  \right ],
\end{eqnarray}
where $a(t)$ is the scale factor, which depends on the cosmic time $t$, and $k=\{0,\pm1\}$ is the spatial curvature parameter. We have taken $c=1$.

\subsection{No matter fields}

We begin by establishing the equations of motion for trace-free Einstein gravity in the absence of matter fields. In this scenario,~\eqref{TFEM} reduces to 
 \begin{eqnarray}\label{TFEG_without_matter}
 R_{\mu\nu} - \frac14 R g_{\mu\nu} =0.
 \end{eqnarray}
Using~\eqref{FRWspacetime}, these equations acquire the form
\begin{eqnarray}\label{TFEG_cosmology}
{\mathcal E} = 0,
\end{eqnarray}
with
\begin{eqnarray}\label{IP_nomatter}
{\mathcal E} := \frac{\ddot a}{a} - \frac{{\dot a}^2}{a^2} - \frac{k}{a^2}= \frac{d}{dt} \left ( \frac{\dot a}{a}\right ) - \frac{k}{a^2},
\end{eqnarray}
where the dot denotes total derivative with respect to the cosmic time $t$. Note that we get just a single equation, not two as in general relativity [see~\eqref{Friedmann} and~\eqref{GR} below].

{\it Constant of motion $\hat{\Lambda}$}. It is not hard to realize that the 
function
\begin{eqnarray}\label{Lambda_cosmology}
\hat{\Lambda} (a, {\dot a}) := 3 \left (\frac{{\dot a}^2}{a^2} + \frac{k}{a^2}  \right )
\end{eqnarray}
is a constant of motion for the system given by equation~\eqref{TFEG_cosmology}, In fact, 
\begin{eqnarray}\label{com_Lambda}
\frac{d\hat{\Lambda}}{dt} = 6 \left ( \frac{\dot a}{a} \right){\mathcal E}
\end{eqnarray}
holds off-shell and hence, on-shell $({\mathcal E} = 0)$, we have $d {\hat{\Lambda}}/dt=0$. Therefore, $\hat{\Lambda}$ remains constant with respect to the cosmic time $t$ provided that the equation of motion~\eqref{TFEG_cosmology} is satisfied, and its value can be calculated from the value of $a$ and its time derivative ${\dot a}$ at any time $t$ using the expression~\eqref{Lambda_cosmology}. It is also worth noting that using~\eqref{Lambda_cosmology}, the trace-free Einstein equation~\eqref{TFEG_cosmology} can be written as
\begin{eqnarray}
\ddot{a} = \frac{1}{3} \hat{\Lambda} a.
\end{eqnarray}

On the other hand, the expression for the constant of motion~\eqref{Lambda_cosmology} and the equation of motion of trace-free Einstein gravity~\eqref{TFEG_cosmology} can be written in more familiar form as
\begin{eqnarray}
- 3 \left (\frac{{\dot a}^2}{a^2} + \frac{k}{a^2}  \right ) + \hat{\Lambda} &=&0,\label{Lambda_cosmology_2}\\
- 2 \frac{\ddot a}{a} - \frac{{\dot a}^2}{a^2} - \frac{k}{a^2} + \hat{\Lambda}  &=&0,\label{tfee_cosmology_rewritten}
\end{eqnarray}
respectively. These expressions closely resemble the equations of motion of general relativity with cosmological constant $\Lambda$,
\begin{eqnarray}\label{GR+Lambda}
R_{\mu\nu} - \frac12 R g_{\mu\nu} + \Lambda g_{\mu\nu} =0.
\end{eqnarray}
Indeed, using~\eqref{FRWspacetime}, the equations~\eqref{GR+Lambda} take the form
\begin{eqnarray}
- 3  \left (\frac{{\dot a}^2}{a^2} + \frac{k}{a^2}  \right ) + \Lambda &=&0, \label{Friedmann}\\
- 2 \frac{\ddot a}{a} - \frac{{\dot a}^2}{a^2} - \frac{k}{a^2} + \Lambda  &=&0. \label{GR}
\end{eqnarray}

By comparing~\eqref{Friedmann} (the first Friedmann equation) with~\eqref{Lambda_cosmology_2}, it becomes natural to refer to the constant of motion $\hat{\Lambda}$ the `cosmological constant'. However, it is important to remark that $\hat{\Lambda}$ and $\Lambda$ have different meanings. In general relativity, $\Lambda$ is a parameter with a specific (though unknown) value that is the same for all of the solutions of~\eqref{GR+Lambda}. In contrast, in trace-free Einstein gravity, $\hat{\Lambda}$ does not appear in~\eqref{TFEG_without_matter} and does not have a specific value. Instead, the value of $\hat{\Lambda}$ depends on the initial data, as given by~\eqref{Lambda_cosmology}.

Finally, let us note that the scalar curvature $R$ for the FRW spacetime acquires the form
\begin{eqnarray}
R = 6 \left ( \frac{\ddot a}{a} + \frac{{\dot a}^2}{a^2} + \frac{k}{a^2} \right ) = 4 \hat{\Lambda} + 6 {\mathcal E}.
\end{eqnarray}
Therefore, on-shell $({\mathcal E} = 0)$, it becomes
\begin{eqnarray}
R = 4 \hat{\Lambda}.
\end{eqnarray}

\subsubsection{Lagrangian formulations}\,

{\bf First formulation}. We note that the equation of motion~\eqref{TFEG_cosmology} is deduced from the action
\begin{eqnarray}\label{cosmology_action}
S[a] = \int^{t_2}_{t_1} {\mathcal L} dt, \quad {\mathcal L} = \frac12 \left( \frac{{\dot a}^2}{a^2} -  \frac{k}{a^2} \right).
\end{eqnarray}
Indeed, using~\eqref{cosmology_action}, we have
\begin{eqnarray}\label{eq_m_tf}
\frac{\partial {\mathcal L}}{\partial a}  - \frac{d}{dt} \left ( \frac{\partial {\mathcal L}}{\partial {\dot a}} \right )  
= - \frac{1}{a} \left (\frac{{\ddot a}}{a} - \frac{{\dot a}^2}{a^2} - \frac{k}{a^2} \right )=0, 
\end{eqnarray}
which clearly yields the equation of motion~\eqref{TFEG_cosmology}. The action~\eqref{cosmology_action} is the first action in the metric formalism reported in literature for trace-free Einstein gravity in the absence of matter fields for the case of the FRW spacetime.   

It is worth noting that the action~\eqref{cosmology_action} is invariant under translation in the cosmic time $t$:
\begin{eqnarray}
t' = t + \alpha, \quad \alpha=\mbox{arbitrary constant parameter}, \nonumber\\
a'(t') = a(t).
\end{eqnarray}
According to the first Noether theorem~\cite{Noether1918}, this implies the existence of a constant of motion associated with this symmetry. By calculating the corresponding conserved quantity, we get precisely the constant of motion~\eqref{Lambda_cosmology}. 
Alternatively, it is important to note that the constant of motion~\eqref{Lambda_cosmology} is directly proportional to the Jacobi integral. Specifically, we have
\begin{eqnarray}
\hat{\Lambda} = 6 \left ( {\dot a} \frac{\partial {\mathcal L}}{\partial {\dot a}} - {\mathcal L} \right ).
\end{eqnarray}

{\bf Second formulation}. We can obtain an alternative Lagrangian formulation from the action~\eqref{cosmology_action} by performing the coordinate transformation $q=\ln{a}$. This  transformation yields 
\begin{eqnarray}\label{cosmology_action2}
S[q] = \int^{t_2}_{t_1} {\mathcal L} dt, \quad {\mathcal L}= \frac12 \left( {\dot q}^2 - k  e^{-2q} \right),
\end{eqnarray}
which leads to the equation of motion
\begin{eqnarray}
\frac{\partial {\mathcal L}}{\partial q}  - \frac{d}{dt} \left ( \frac{\partial {\mathcal L}}{\partial {\dot q}} \right )  
= -{\ddot q} + k e^{-2q}=0.
\end{eqnarray}
A key feature of this Lagrangian formulation is that it naturally leads to a canonical formulation involving the expansion rate $H$ as the momentum canonically conjugate to $q$ (see below).

Obviously, 
\begin{eqnarray}
\hat{\Lambda} (q,{\dot q}) &=& 6 \left ( {\dot q} \frac{\partial {\mathcal L}}{\partial {\dot q}} - {\mathcal L} \right ) \nonumber\\
&=& 3 \left ( {\dot q}^2 + k e^{-2q}\right )
\end{eqnarray}
is the constant of motion~\eqref{Lambda_cosmology} but expressed in terms of $q$ and ${\dot q}$.

\subsubsection{Hamiltonian formulations}\,

{\bf First canonical formulation}. To derive the canonical formulation from the Lagrangian formulation~\eqref{cosmology_action}, we define the momentum $\pi$, canonically conjugate to $a$, in the usual way 
\begin{eqnarray}
\pi := \frac{\partial {\mathcal L}}{\partial {\dot a}} = \frac{\dot a}{a^2}.
\end{eqnarray}
The action~\eqref{cosmology_action} acquires the canonical form
\begin{eqnarray}\label{CF}
S[a, \pi] = \int^{t_2}_{t_1} \left ( \pi {\dot a} - {\mathcal H} \right ) dt, \nonumber\\
{\mathcal H} = \frac12 \left( a^2 \pi^2 + \frac{k}{a^2} \right), 
\end{eqnarray}
which gives the Hamilton equations
\begin{eqnarray}\label{Hamilton_eqs}
    {\dot a} =& \frac{\partial {\mathcal H}}{\partial \pi} = a^2 \pi,\\
    {\dot \pi} =& - \frac{\partial {\mathcal H}}{\partial a} = - a \pi^2 + \frac{k}{a^3}.
\end{eqnarray}

Notably, in terms of the canonical coordinates, the constant of motion $\hat{\Lambda}$ can be written as
\begin{eqnarray}
\hat{\Lambda} = 6 {\mathcal H}.
\end{eqnarray}

{\bf First noncanonical formulation}. Instead of using the canonical variables $(a,\pi)$, we can alternatively label locally the points of the phase space with $a$ and the expansion rate $H={\dot a}/a$. Therefore, we have
\begin{eqnarray}\label{NC1}
{\dot a} = a H.
\end{eqnarray}
This approach is advantageous not only due to $H$ is a key variable employed in cosmology, but also because it arises naturally from equation of motion~\eqref{TFEG_cosmology}. In fact, using~\eqref{NC1}, equation~\eqref{TFEG_cosmology} acquires the simple form
\begin{eqnarray}\label{NC2}
{\dot H} = \frac{k}{a^2}.
\end{eqnarray}

Note that~\eqref{NC1} and~\eqref{NC2} can be written in Hamiltonian form as
\begin{eqnarray}
{\dot x}^a = \omega^{ab} \frac{\partial {\mathcal H}}{\partial x^b},
\end{eqnarray}
where  $(x^a)=(x^1,x^2)=(a,H)$ and
\begin{eqnarray}
\omega^{12} =& a = - \omega^{21}, \quad \omega^{11}=0= \omega^{22}, \nonumber\\
{\mathcal H} =& \frac12 \left( H^2 + \frac{k}{a^2} \right).
\end{eqnarray}
This means that $(a,H)$ are not canonical coordinates. Actually, they satisfy
\begin{eqnarray}
\{ a , H \} =a,
\end{eqnarray}
which implies that the symplectic structure $\omega$ on the phase space is locally given by
\begin{eqnarray}
\omega = \frac{1}{a} d H \wedge da,
\end{eqnarray}
where $\wedge$ is the wedge product.

Equations~\eqref{NC1} and~\eqref{NC2} come from the action 
\begin{eqnarray}\label{NCHF}
S [a , H] = \int^{t_2}_{t_1} \left [ \left ( \frac{H}{a}\right ) {\dot a} - \frac12 \left ( H^2 + \frac{k}{a^2}\right ) \right ] dt.
\end{eqnarray}
Note incidentally that by using Darboux's theorem~\cite{arnold2006}, we can obtain the canonical formulation~\eqref{CF} from~\eqref{NCHF}. The Darboux map that achieves this transformation is
\begin{eqnarray}
(a, H) \mapsto (a,\pi)=\left(a, \frac{H}{a} \right).
\end{eqnarray}

{\bf Second canonical formulation}. We can also derive another canonical formulation from the Hamiltonian formulation~\eqref{NCHF} by using the Darboux map 
\begin{eqnarray}
(a, H) \mapsto (q,H)=(\ln{a}, H).
\end{eqnarray}
It is straightforward to verify that $(q,H)$ are indeed canonical coordinates, as they satisfy
\begin{eqnarray}
\{ q, H \} = 1.  
\end{eqnarray}
In terms of these variables, the action~\eqref{NCHF} takes the canonical form
\begin{eqnarray}\label{CF2}
S[q,H] = \int^{t_2}_{t_1} \Big [ H \dot{q} - {\mathcal H} \Big ] d t, \nonumber\\
{\mathcal H} = \frac12 \left ( H^2 + k e^{-2q} \right ),
\end{eqnarray}
which leads to the Hamilton equations
\begin{eqnarray}
{\dot q} = \frac{\partial {\mathcal H}}{\partial H} = H, \\
{\dot H} = - \frac{\partial {\mathcal H}}{\partial q} = k e^{-2 q}.
\end{eqnarray}

Note that the canonical formulation~\eqref{CF2} is related to the canonical formulation~\eqref{CF} through the canonical transformation
\begin{eqnarray}
q = \ln{a}, \\
H = a \pi.
\end{eqnarray}
Furthermore, the canonical formulation~\eqref{CF2} can also be obtained directly by performing the Legendre transform of the Lagrangian action~\eqref{cosmology_action2}.

{\bf Second noncanonical formulation}. By performing the coordinate transformation $a=e^q$ in the canonical formulation~\eqref{CF}, we get the action
\begin{eqnarray}\label{NCHF2}
S[q,\pi] = \int^{t_2}_{t_1} \left [ \pi e^{q} {\dot q} - {\mathcal H}\right ] dt, \nonumber \\
{\mathcal H}= \frac12 \left ( e^{2q} \pi^2 + k e^{-2q} \right ).
\end{eqnarray}
Note that the phase space variables $(q,\pi)$ in this formulation are non canonical, as they satisfy
\begin{eqnarray}
\{ q, \pi \}= e^{-q}.
\end{eqnarray}
Lastly, it is worth highlighting that the relevance of the Hamiltonian formulation~\eqref{NCHF2} lies in the fact that it leads to the canonical formulation~\eqref{CF2} by using Darboux's theorem~\cite{arnold2006}. This can be accomplished by using the Darboux map 
\begin{eqnarray}
(q,\pi) \rightarrow (q,H)= (q, \pi e^q).
\end{eqnarray}

\begin{figure}
\begin{center}
\begin{tikzpicture}[scale=1.5]

\draw[gray, thick,<-] (-2.2,1.75) -- (-0.1,1.75);
\draw[gray, thick,<->] (-2.2,-0.5) -- (-0.1,-0.5);
\draw[gray, thick,->] (-2.2,-2.7) -- (-0.1,-2.7);

\draw[gray, thick,->] (-2.75,1.37) -- (-2.75,-0.15);
\draw[gray, thick,->] (0.3,1.37) -- (0.3,-0.15);
\draw[gray, thick,<->] (0.6,1.37) -- (0.6,-0.15);

\draw[gray, thick,<->] (-2.95,-2.35) -- (-2.95,-0.87);
\draw[gray, thick,->] (-2.65,-2.35) -- (-2.65,-0.87);
\draw[gray, thick,->] (0.45,-2.35) -- (0.45,-0.87);

\draw[gray, thick,<-] (-2.35,-0.2) -- (0,1.5);
\draw[gray, thick,->] (-2.35,-2.45) -- (0,-0.8);

\draw[gray, thick,->] (-3.27,1.75) --(-3.9,1.75) -- (-3.9,-3.3) -- (0.45,-3.3) -- (0.45,-3.05); 
 
\draw[black] (-3,1.75)  node[anchor=west]{$S[a]$};
\draw[black] (0,-.5)  node[anchor=west]{$S[q,H]$};
\draw[black] (-3.2,-0.5)  node[anchor=west]{$S[a,\pi]$};
\draw[black] (0,1.75)  node[anchor=west]{$S[a,H]$};
\draw[black] (-3.2,-2.7)  node[anchor=west]{$S[q,\pi]$};
\draw[black] (0.1,-2.7)  node[anchor=west]{$S[q]$};

\draw[black] (-1.1,2)  node[anchor=mid,rotate=0,font=\fontsize{8}{0}\selectfont]{$H=\frac{\dot{a}}{a}$};

\draw[black] (-3.10,0.65)  node[anchor=mid,rotate=90,font=\fontsize{8}{0}\selectfont]{Legendre}; 
\draw[black] (-2.9,0.65)  node[anchor=mid,rotate=90,font=\fontsize{8}{0}\selectfont]{transformation};

\draw[black] (0.65,-1.66)  node[anchor=mid,rotate=90,font=\fontsize{8}{0}\selectfont]{Legendre}; 
\draw[black] (0.87,-1.68)  node[anchor=mid,rotate=90,font=\fontsize{8}{0}\selectfont]{transformation};

\draw[black] (-1.3,0.75)  node[anchor=mid,rotate=37,font=\fontsize{8}{0}\selectfont]{Darboux map}; 
\draw[black] (-1.3,-1.52)  node[anchor=mid,,rotate=37,font=\fontsize{8}{0}\selectfont]{Darboux map};

\draw[black] (-1.1,-0.15)  node[anchor=mid,font=\fontsize{8}{0}\selectfont]{Canonical};
\draw[black] (-1.1,-0.33)  node[anchor=mid,font=\fontsize{8}{0}\selectfont]{transformation};

\draw[black] (0.15,0.65)  node[anchor=mid,rotate=90,font=\fontsize{8}{0}\selectfont]{Darboux map}; 

\draw[black] (-2.47,-1.65)  node[anchor=mid,rotate=90,font=\fontsize{8}{0}\selectfont]{Darboux map}; 

\draw[black] (-1.7,-3.5)  node[anchor=mid,rotate=0,font=\fontsize{8}{0}\selectfont]{Coordinate transformation};

\draw[black] (0.62,1.2)  node[anchor=west,font=\fontsize{8}{0}\selectfont]{$a=e^q$}; 
\draw[black] (0.62,0.12)  node[anchor=west,font=\fontsize{8}{0}\selectfont]{$q=\ln{a}$}; 

\draw[black] (-2.95,-1.10)  node[anchor=east,font=\fontsize{8}{0}\selectfont]{$a=e^q$}; 
\draw[black] (-2.97,-2.17)  node[anchor=east,font=\fontsize{8}{0}\selectfont]{$q=\ln{a}$}; 

\draw[black] (-0.68,-2.9)  node[anchor=east,font=\fontsize{8}{0}\selectfont]{$\pi=e^{-q}\dot{q}$};

\end{tikzpicture}
\end{center}
  \caption{Lagrangian and Hamiltonian formulations for trace-free Einstein gravity  with no matter fields in the FRW spacetime.}
    \label{fig:actions}
\end{figure}
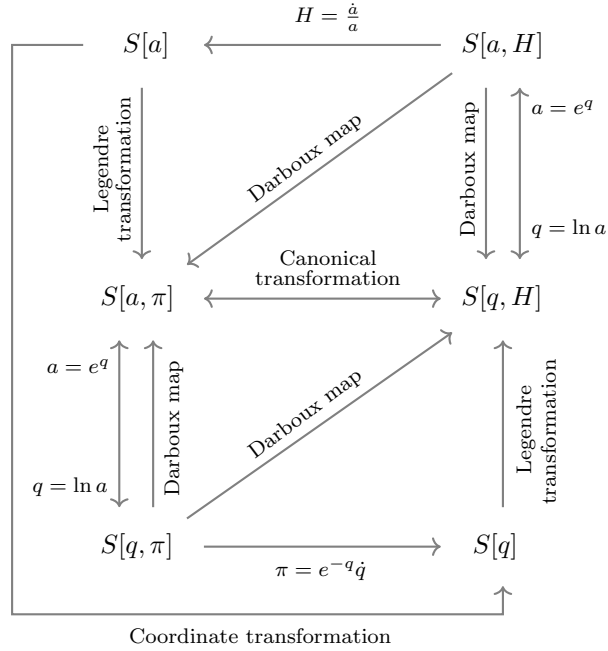

To conclude this section, figure~\ref{fig:actions} provides a summary of the Lagrangian and Hamiltonian formulations obtained for trace-free Einstein gravity in the absence of matter fields in the FRW spacetime.

\subsection{Matter fields}

\subsubsection{Inflaton and phantom}\,

We begin by considering the coupling of inflaton $\Phi$ and phantom $\varphi$ fields to trace-free Einstein gravity. The equations of motion~\eqref{TFEM} take the form\footnote{The energy-momentum tensor of $\Phi$ and $\varphi$ is $T_{\mu\nu}= \partial_{\mu} \Phi \partial_{\nu} \Phi - \partial_{\mu} \varphi \partial_{\nu} \varphi  - g_{\mu\nu} \left ( \frac12 g^{\alpha\beta} \partial_{\alpha} \Phi \partial_{\beta} \Phi - \frac12 g^{\alpha\beta} \partial_{\alpha} \varphi \partial_{\beta} \varphi + U(\Phi,\varphi) \right )$.}
\begin{eqnarray}\label{IP1}
{\mathcal E} = 0,
\end{eqnarray}
with
\begin{eqnarray}\label{IP1*}
{\mathcal E} := \frac{\ddot a}{a} - \frac{{\dot a}^2}{a^2} - \frac{k}{a^2} + \frac{\kappa^2}{2} \left (  {\dot \Phi}^2 -  {\dot \varphi}^2 \right ).
\end{eqnarray}
Additionally, the equations of motion for the inflaton $\Phi$ and phantom $\varphi$ fields are given by~\cite{Ellis_2011,Ellis_2014}
\begin{eqnarray}
{\mathcal E}_{\Phi} = 0, \label{IP2}\\
{\mathcal E}_{\varphi} = 0, \label{IP3}
\end{eqnarray}
respectively, with
\begin{eqnarray}
{\mathcal E}_{\Phi} := {\ddot \Phi} + 3 \frac{\dot a}{a} {\dot \Phi} + \frac{\partial U}{\partial \Phi}, \label{EL_I}\\
{\mathcal E}_{\varphi} := {\ddot \varphi} + 3 \frac{\dot a}{a} {\dot \varphi} - \frac{\partial U}{\partial \varphi}, \label{EL_P}
\end{eqnarray}
where $U \equiv U(\Phi,\varphi)$ is a potential that depends only on $\Phi$ and $\varphi$. Note that $U$ does not appear in~\eqref{IP1*}. 

{\it Constant of motion $\hat{\Lambda}$}. It can be verified that the expression 
\begin{eqnarray}\label{cmIP}
\hat{\Lambda} (a,\Phi,\varphi, {\dot a}, \dot{\Phi}, \dot{\varphi}) := 3 \left ( \frac{{\dot a}^2}{a^2} + \frac{k}{a^2}  \right )  - \kappa^2 \left [ \frac12 {\dot \Phi}^2 - \frac12 {\dot{\varphi}}^2 + U \right ]
\end{eqnarray}
is a constant of motion for the system given by~\eqref{IP1},~\eqref{IP2}, and~\eqref{IP3}. In fact, we have the off-shell identity
\begin{eqnarray}
\frac{d \hat{\Lambda}}{dt} = 6 \left ( \frac{\dot a}{a} \right ) {\mathcal E} - \kappa^2 {\dot \Phi} \, {\mathcal E}_{\Phi} + \kappa^2 {\dot\varphi} \, {\mathcal E}_{\varphi},
\end{eqnarray}
and therefore, on-shell (${\mathcal E}=0$, ${\mathcal E}_{\Phi}=0$, ${\mathcal E}_{\varphi}=0$), we get $d {\hat{\Lambda}}/dt=0$. From this is clear that $\hat{\Lambda}$ remains constant with respect to the cosmic time $t$, provided the equations of motion~\eqref{IP1},~\eqref{IP2}, and~\eqref{IP3} are satisfied, and its value depends on the initial data according to~\eqref{cmIP}. Also, it is not hard to see that, using~\eqref{cmIP}, the equation of motion~\eqref{IP1} can be written as
\begin{eqnarray}\label{aceleration}
\ddot{a} =\frac13 \left [ \hat{\Lambda} +\kappa^2 \left ( U - {\dot \Phi}^2 + {\dot\varphi}^2 \right ) \right ] a.
\end{eqnarray}

On the other hand, it is worth noticing that more familiar expressions for the constant of motion~\eqref{cmIP} and the equation of motion~\eqref{IP1} are
\begin{eqnarray}
- 3 \left ( \frac{{\dot a}^2}{a^2} + \frac{k}{a^2}  \right ) + \hat{\Lambda} = \kappa^2 \left ( - \frac12 {\dot \Phi}^2 + \frac12 {\dot{\varphi}}^2 - U \right ), \label{cmIP_rewritten}\\
- 2 \frac{\ddot a}{a} - \frac{{\dot a}^2}{a^2} - \frac{k}{a^2} + \hat{\Lambda}  = \kappa^2 \left ( \frac12 {\dot \Phi}^2 - \frac12 {\dot \varphi}^2 - U \right ), \label{IP1_rewritten}
\end{eqnarray}
respectively. The expressions~\eqref{cmIP_rewritten} and~\eqref{IP1_rewritten} resemble the equations of general relativity with cosmological constant $\Lambda$ and an energy-momentum tensor for $\Phi$ and $\varphi$,
\begin{eqnarray}
R_{\mu\nu} - \frac12 R g_{\mu\nu} + \Lambda g_{\mu\nu} = \kappa^2 T_{\mu\nu},
\end{eqnarray}
calculated with the FRW metric~\eqref{FRWspacetime}. That is why the constant of motion $\hat{\Lambda}$ is referred to as the cosmological constant. However, it is important to remark that, unlike the cosmological constant $\Lambda$ in general relativity, $\hat{\Lambda}$ is not a parameter with a fixed value.

Before closing this section, we would like to point out that the scalar curvature $R$ and the trace $T$ of the energy-momentum tensor of $\Phi$ and $\varphi$ satisfy the off-shell identity
\begin{eqnarray}
R + \kappa^2 T= 4 \hat{\Lambda} + 6 {\mathcal E},
\end{eqnarray}
which, on-shell (${\mathcal E}=0$), reduces to
\begin{eqnarray}
R + \kappa^2 T = 4 \hat{\Lambda}.
\end{eqnarray}

{\it Additional constant of motion ${\mathcal C}_1$}. It is straightforward to verify that the expression 
\begin{eqnarray}\label{angular}
{\mathcal C}_1 := a^3 \left ( \Phi {\dot \varphi} - \varphi {\dot \Phi} \right )
\end{eqnarray}
is a constant of motion, provided the potential $U$ is an arbitrary function of the $SO(1,1)$ invariant $\Phi^2 - \varphi^2$, i.e., $U=U(\Phi^2 - \varphi^2)$. Note that this $SO(1,1)$ invariant also exists in the space of inflaton and phantom fields coupled to general relativity~\cite{Abdel_Montesinos_Matos}. In fact, under the assumption that $U=U(\Phi^2 - \varphi^2)$, we have the off-shell identity
\begin{eqnarray}\label{Id_cml}
\frac{d {\mathcal C}_1}{dt} = a^3 \left(\Phi \, {\mathcal E}_{\varphi} - \varphi \, {\mathcal E}_{\Phi}\right),
\end{eqnarray}
and therefore, on-shell (${\mathcal E}_{\Phi}=0$ and ${\mathcal E}_{\varphi}=0$), we get $ d {\mathcal C}_1 / dt =0$. This clearly shows that ${\mathcal C}_1$ remains constant with respect to the cosmic time $t$, provided that ${\mathcal E}_{\Phi}=0$ and ${\mathcal E}_{\varphi}=0$ hold. It is interesting to note that ${\mathcal E}$ does not appear in the identity~\eqref{Id_cml}, meaning that ${\mathcal C}_1$ is also a constant of motion for any other geometric theory coupled to $\Phi$ and $\varphi$ in the cosmological scenario described by the FRW metric. This includes, for instance, the cosmology of general relativity with $\Phi$ and $\varphi$ as matter fields, as we pointed out above.

{\it Further constant of motion if $U=U_0$, where $U_0$ is a constant}.  Under this assumption, the expressions~\eqref{EL_I} and~\eqref{EL_P} become
\begin{eqnarray}
{\mathcal E}_{\Phi} & =& {\ddot \Phi} + 3 \frac{\dot a}{a} {\dot \Phi}, \\
{\mathcal E}_{\varphi} & =& {\ddot \varphi} + 3 \frac{\dot a}{a} {\dot \varphi}.
\end{eqnarray}
Therefore, 
\begin{eqnarray}\label{momento}
{\mathcal C}_2 := a^6 \left [ ({\dot \Phi})^2 - ({\dot \varphi})^2 \right ]
\end{eqnarray}
is also a constant of motion because the off-shell identity 
\begin{eqnarray}
\frac{d{\mathcal C}_2}{dt} = 2 a^6 \left ( {\dot \Phi} {\mathcal E}_{\Phi} - {\dot \varphi} {\mathcal E}_{\varphi} \right )
\end{eqnarray}
holds. 

{\it Equation of the orbit if $U=U_0$}. In this particular case, it is easy to deduce the equation of the orbit. In fact,
\begin{eqnarray}
{\ddot \Phi} + 3 \frac{\dot a}{a} {\dot \Phi} =0
\end{eqnarray}
amounts to
\begin{eqnarray}\label{almost_orbit}
\frac{\ddot a}{a} \frac{d \Phi}{d a} + \frac{({\dot a})^2}{a} \frac{d^2 \Phi}{d a^2} + 3 \frac{({\dot a})^2}{a^2} \frac{d \Phi}{d a} =0.
\end{eqnarray}

On the other hand, expressions~\eqref{aceleration} and~\eqref{angular} imply
\begin{eqnarray}
\frac{\ddot a}{a} &=& \frac13 \left ( \hat{\Lambda} + \kappa^2 U_0 - \kappa^2 {\mathcal C}_2 a^{-6} \right ), \\
{\dot a} &=& {\mathcal C}_1 a^{-3} \left ( \Phi \frac{d \varphi}{da} - \varphi \frac{d \Phi}{d a} \right )^{-1},
\end{eqnarray}
respectively. 

Substituting these two expressions in~\eqref{almost_orbit}, we get the equation 
\begin{eqnarray}\label{re_1}
 \frac13 \left ( \hat{\Lambda} + \kappa^2 U_0 - \kappa^2 {\mathcal C}_2 a^{-6} \right ) \frac{d \Phi}{d a}  + {\mathcal C}_1^2 a^{-7} \left ( \Phi \frac{d \varphi}{da} - \varphi \frac{d \Phi}{d a} \right )^{-2} \frac{d^2 \Phi}{d a^2} \nonumber\\ + 3 {\mathcal C}_1^2 a^{-8} \left ( \Phi \frac{d \varphi}{da} - \varphi \frac{d \Phi}{d a} \right )^{-2} \frac{d \Phi}{d a} =0.
\end{eqnarray}
Similarly,
\begin{eqnarray}
{\ddot \varphi} + 3 \frac{\dot a}{a} {\dot \varphi} =0
\end{eqnarray}
leads to
\begin{eqnarray}\label{re_2}
 \frac13 \left ( \hat{\Lambda} + \kappa^2 U_0 - \kappa^2 {\mathcal C}_2 a^{-6} \right ) \frac{d \varphi}{d a}  + {\mathcal C}_1^2 a^{-7} \left [ \Phi \frac{d \varphi}{da} - \varphi \frac{d \Phi}{d a} \right ]^{-2} \frac{d^2 \varphi}{d a^2} \nonumber\\
 + 3 {\mathcal C}_1^2 a^{-8} \left [ \Phi \frac{d \varphi}{da} - \varphi \frac{d \Phi}{d a} \right ]^{-2} \frac{d \varphi}{d a} =0.
\end{eqnarray}
Expressions~\eqref{re_1} and~\eqref{re_2} give $\Phi$ and $\varphi$ parameterized in terms of $a$.

However, note that 
\begin{eqnarray}\label{re_3}
\fl \hat{\Lambda} = 3 {\mathcal C}_1^2 a^{-8} \left [ \Phi \left ( \frac{d \varphi}{d a} \right ) - \varphi \left ( \frac{d \Phi}{da} \right ) \right]^{-2}  + 3 k a^{-2} - \frac12 \kappa^2 {\mathcal C}_2 a^{-6} - \kappa^2 U_0
\end{eqnarray}
also holds, which comes from~\eqref{cmIP},~\eqref{angular}, and~\eqref{momento}. Alternatively, this expression comes from~\eqref{IP1} and~\eqref{IP1*}.

Therefore, using~\eqref{re_3}, the expressions~\eqref{re_1} and~\eqref{re_2} acquire the simpler form
\begin{eqnarray}
&& a \left ( \hat{\Lambda} + \kappa^2 U_0 - 3k a^{-2} + \frac12 \kappa^2 {\mathcal C}_2 a^{-6} \right ) \frac{d^2 \Phi}{d a^2} \nonumber\\
&& + \left ( 4 \hat{\Lambda} + 4 \kappa^2 U_0 - 9 k a^{-2} + \frac12 \kappa^2 {\mathcal C}_2 a^{-6} \right ) \frac{d \Phi}{d a} =0, \\
&& a \left ( \hat{\Lambda} + \kappa^2 U_0 - 3k a^{-2} + \frac12 \kappa^2 {\mathcal C}_2 a^{-6} \right ) \frac{d^2 \varphi}{d a^2} \nonumber\\
&& + \left ( 4 \hat{\Lambda} + 4 \kappa^2 U_0 - 9 k a^{-2} + \frac12 \kappa^2 {\mathcal C}_2 a^{-6} \right ) \frac{d \varphi}{d a} =0. 
\end{eqnarray}

These expressions admit, of course, another interpretation: $a$ plays the role of a physical clock, and $\Phi$ and $\varphi$ `evolve' with respect to it. Therefore, the previous expressions display the relational evolution of the variables $a$, $\Phi$, and $\varphi$. See~\cite{PhysRevD.60.044009,PhysRevD.65.124013} (and references therein) for more on the relational evolution.  

{\it On the Lagrangian formulation}. 
We might wonder about the Lagrangian ${\mathcal L}$ that allows us the construction of an action $S[a,\Phi,\varphi]$ whose variation leads to the equations~\eqref{IP1},~\eqref{IP2}, and~\eqref{IP3}. However, it can be shown that such a Lagrangian does not exist if the corresponding  action is required to be invariant under the rigid translation in the cosmic time $t$
\begin{eqnarray}\label{rtt}
t' &=& t + \alpha, \quad \alpha=\mbox{arbitrary constant parameter}, \nonumber\\
a'(t') &=& a(t), \quad \Phi' (t') = \Phi (t), \quad \varphi' (t') = \varphi (t).
\end{eqnarray}
The reason is that this symmetry requirement is incompatible with the Helmholtz conditions~\cite{Douglas_paper}. Consequently, the constant of motion~\eqref{cmIP} cannot be obtained through the first Noether theorem~\cite{Noether1918} using the rigid translation in the time $t$~\eqref{rtt}. 

\subsubsection{Perfect fluid.} \ 

We now focus our attention on trace-free Einstein gravity in the presence of a perfect fluid. The equations of motion~\eqref{TFEM} take the form
\begin{eqnarray}\label{TF+fluid}
   {\mathcal E}=0, 
\end{eqnarray}
with
\begin{eqnarray}
    {\mathcal E} := \frac{\ddot a}{a} - \frac{{\dot a}^2}{a^2} - \frac{k}{a^2} + \kappa^2 \frac12 (p+\rho),
\end{eqnarray}
while the equation of motion of the perfect fluid is 
\begin{eqnarray}\label{TFcosmology+fluid}
 {\mathcal E}_{\rho}=0,
\end{eqnarray}
with 
\begin{eqnarray}\label{dervar5}
{\mathcal E}_{\rho} := {\dot \rho} + 3 \left ( \frac{{\dot a}}{a} \right ) \left ( p + \rho \right ),
\end{eqnarray}
where $p$ and $\rho$ are the pressure and energy density, respectively.

{\it Constant of motion $\hat{\Lambda}$}. It can be checked that the expression
\begin{eqnarray}\label{cmfluid}
\hat{\Lambda} (a, {\dot a}, \rho) = 3 \Big ( \frac{{\dot a}^2}{a^2} + \frac{k}{a^2} \Big ) - \kappa^2 \rho
\end{eqnarray}
is a constant of motion for the system described by~\eqref{TF+fluid} and ~\eqref{TFcosmology+fluid}, regardless of the equation of state for the perfect fluid. In fact, the off-shell identity 
\begin{eqnarray}
\frac{d \hat{\Lambda}}{dt} = 6 \left ( \frac{\dot a}{a} \right ) {\mathcal E} - \kappa^2 {\mathcal E}_{\rho}
\end{eqnarray}
holds, and then on-shell (${\mathcal E}=0$, ${\mathcal E}_{\rho}=0$), we find that $d\hat{\Lambda} / dt = 0$. Clearly,  $\hat{\Lambda}$ remains constant with respect to the cosmic time $t$, provided that the equations of motion~\eqref{TF+fluid} and~\eqref{TFcosmology+fluid} are satisfied. 

Note that more familiar expressions for the constant of motion~\eqref{cmfluid} and the equation of motion~\eqref{TF+fluid} are
\begin{eqnarray}
- 3 \Big ( \frac{{\dot a}^2}{a^2} + \frac{k}{a^2} \Big ) + \hat{\Lambda} = - \kappa^2 \rho, \label{GR+fluid1} \\
- 2 \frac{\ddot a}{a} - \frac{{\dot a}^2}{a^2} - \frac{k}{a^2} + \hat{\Lambda}  = \kappa^2 p, \label{GR+fluid2}
\end{eqnarray}
respectively. These equations resemble Einstein's equations for general relativity with cosmological constant $\Lambda$ in the FRW spacetime. It is also worth noting that using using~\eqref{cmfluid}, the equation of motion~\eqref{TF+fluid} can be written as
\begin{eqnarray}
\ddot{a} = \frac13 \left [ \hat{\Lambda} - \frac12 \kappa^2 \left (\rho + 3 p \right )\right ] a. 
\end{eqnarray}
Moreover, note that the scalar curvature $R$ and the trace $T$ of the energy momentum tensor for the perfect fluid fulfill the off-shell identity
\begin{eqnarray}
R + \kappa^2 T= 4 \hat{\Lambda} + 6 {\mathcal E},
\end{eqnarray}
which, on-shell (${\mathcal E}=0$), reduces to
\begin{eqnarray}
\left ( R + \kappa^2 T \right )= 4 \hat{\Lambda}.
\end{eqnarray}

We close this section by making two comments: 1) Note that if we consider a vacuum energy with equation of state $p_{\Lambda}= - \rho_{\Lambda}$ with $\rho_{\Lambda}$ a given constant, then the expression~\eqref{cmfluid} clearly shows that the constant of motion $\hat{\Lambda}$ is not the vacuum energy. 2) The various matter couplings considered illustrate the general fact that the expression of the constant of motion $\hat{\Lambda}$ depends on the type of matter fields coupled.

\section{Trace-free Einstein cosmology in the conformal time}\label{ct_evolution}

In this section, we come back to the theory in the absence of matter fields and use the conformal time $\eta$ as the evolution parameter instead of the cosmic time $t$ (for an operational definition of the conformal time, see~\cite{Chiu:1964}). The conformal time $\eta$ is defined by the relation
\begin{eqnarray}
d \eta= \frac{dt}{a}. 
\end{eqnarray}
One advantage of using the conformal time $\eta$ is that the cosmological equations discussed in the previous section become simpler than in the cosmic time $t$. However, this is not enough, a change of variable is required. 

We begin by expressing the equation of motion~\eqref{TFEG_cosmology} in terms of $\eta$. By doing so, equation~\eqref{TFEG_cosmology} becomes
\begin{eqnarray}\label{eq_c_time}
\frac{a''}{a^3} - 2\frac{(a')^2}{a^4} - \frac{k}{a^2} =0,
\end{eqnarray}
where a prime denotes the total derivative with respect to $\eta$. At first glance, equation~\eqref{eq_c_time} looks more complicated than~\eqref{TFEG_cosmology}. However, by performing the change of variable
\begin{eqnarray}\label{conf_defQ}
Q :=- \frac{1}{a},
\end{eqnarray}
the equation of motion~\eqref{eq_c_time} acquires the simple form
\begin{eqnarray}\label{eq_ct}
\frac{d^2 Q}{d \eta^2} + k Q=0. 
\end{eqnarray}

Since $a\in (0,\infty)$, it follows that $Q\in (-\infty,0)$. Therefore, in the absence of matter fields, the classical dynamics of trace-free Einstein cosmology is as follows: for $k=0$, the system behaves like a free particle; for $k=1$, it corresponds to a harmonic oscillator; and for $k=-1$, it represents a repulsive oscillator. Clearly, in each of these cases, the dynamics is restricted to the negative real half-line ($Q<0$). The general solution of~\eqref{eq_ct} is
\begin{subequations}\label{conf_solutions}
\begin{eqnarray}
Q(\eta) &=& c_1 \eta + c_2, \quad k=0,\\
Q (\eta) &=& c_1 \cos{\eta} + c_2 \sin{\eta}  , \quad k=1,\\
Q(\eta) &=& c_1 e^{\eta} + c_2 e^{-\eta}, \quad k=-1,
\end{eqnarray}
\end{subequations}
where $c_1$ and $c_2$ are integration constants. 

{\it Constant of motion $\hat{\Lambda}$}. In terms of $Q$ and the conformal time $\eta$, the constant of motion~\eqref{Lambda_cosmology} acquires the form
\begin{eqnarray}
\hat{\Lambda} &=& 3 \left [ (Q')^2 + k Q^2 \right ].
\end{eqnarray}
Using~\eqref{conf_solutions}, it becomes
\begin{subequations}
  \begin{eqnarray}
\hat{\Lambda} &=& 3 (c_1)^2, \quad k=0,\\
\hat{\Lambda} &=& 3 \left [ (c_1)^2 + (c_2)^2 \right ], \quad k=1,\\
\hat{\Lambda} &=& -12 c_1 c_2, \quad k=-1.
\end{eqnarray}  
\end{subequations}
This implies that $\hat{\Lambda} \in [0,\infty)$ when $k=0$ or $k=1$. In particular, $\hat{\Lambda}=0$ corresponds to the trivial solutions: $c_1=0$ for $k=0$ and, $c_1=0=c_2$ for $k=1$. Similarly, for $k=-1$, we have $\hat{\Lambda} \in (-\infty, \infty)$, with $\hat{\Lambda}=0$ corresponding to $c_1=0$ or $c_2=0$. 

In what follows we provide Lagrangian formulations for the equations of motion~\eqref{eq_c_time} and~\eqref{eq_ct}.

\subsection{Lagrangian formulations}

{\it First formulation}. It can be verified that the equation of motion~\eqref{eq_c_time} comes from the action
\begin{eqnarray}\label{conf_action1}
S[a] = \int^{\eta_2}_{\eta_1} {\mathcal L} d \eta, \quad {\mathcal L} = \frac12 \left [ \frac{(a')^2}{a^4} - \frac{k}{a^2} \right ].
\end{eqnarray}
In fact, using~\eqref{conf_action1}, we have
\begin{eqnarray}
\frac{\partial {\mathcal L}}{\partial a}  - \frac{d}{d\eta} \left ( \frac{\partial {\mathcal L}}{\partial {a'}} \right )  
= \frac{1}{a} \left (\frac{{a''}}{a^3} - 2 \frac{{(a')}^2}{a^4} - \frac{k}{a^2} \right )=0, 
\end{eqnarray}
which directly leads the equation of motion~\eqref{eq_c_time}.

{\it Second formulation}. It is evident that the equation of motion~\eqref{eq_ct} can be derived from the action
\begin{eqnarray}\label{conf_action2}
S[Q]= \int^{\eta_2}_{\eta_1} {\mathcal L} d\eta, \quad {\mathcal L}= \frac12 (Q')^2 - \frac{k}{2} Q^2.
\end{eqnarray}
See~\cite{Lutzky_1978} for the symmetries of the harmonic oscillator and~\cite{Leach_1980} for the symmetries of the repulsive oscillator. 

\subsection{Hamiltonian formulation}

To obtain the canonical formulation from the Lagrangian formulation~\eqref{conf_action2}, we define the momentum $P$, conjugate to $Q$, as 
\begin{eqnarray}
P := \frac{\partial {\mathcal L}}{\partial Q'} = Q'.
\end{eqnarray}
The action~\eqref{conf_action2} then takes the canonical form
\begin{eqnarray}\label{conf_Hamil}
S[Q,P] = \int^{\eta_2}_{\eta_1} \left [ P Q' - {\mathcal H} \right ] d \eta, \quad {\mathcal H}= \frac12 P^2 + \frac{k}{2} Q^2, \
\end{eqnarray}
which leads to the Hamilton equations
\begin{eqnarray}
Q' &=& P, \\
P' &=& - k Q.
\end{eqnarray}
Note that the momentum $P$ is actually the expansion rate $H$. Specifically, we have
\begin{eqnarray}\label{exp_rate}
P = Q' = a \frac{d}{dt} \left ( - \frac{1}{a}\right ) = \frac{{\dot a}}{a} =H.
\end{eqnarray}

Moreover, it can be verified that the constant of motion $\hat{\Lambda}$ is proportional to the Hamiltonian when expressed in terms of the phase space variables. Specifically, we have
\begin{eqnarray}\label{conf_constant}
\hat{\Lambda} = 6 {\mathcal H}.
\end{eqnarray}

{\it Observation}. It is worth noting that instead of using~\eqref{conf_defQ}, we can also introduce the variable
\begin{eqnarray}
X := \frac{1}{a}.
\end{eqnarray}
In this case, the equation of motion~\eqref{eq_c_time} reads
\begin{eqnarray}\label{eq_c_t2}
\frac{d^2 X}{d \eta^2} + k X =0. 
\end{eqnarray}
Since $a\in (0,\infty)$, we have $X \in (0,\infty)$. Note that the general solution of~\eqref{eq_c_t2} is the same as that of~\eqref{eq_ct}. Of course, in this case, we can also obtain the corresponding Lagrangian and Hamiltonian formulations, along the lines as before.

{\it Cyclic evolution}. In the case $k=1$, spacetime exhibits a cyclic evolution in the conformal time $\eta$ by matching the solutions $Q(\eta)$ and $X(\eta)$. 

Throughout this cyclic evolution, $\hat{\Lambda}$ remains positive and finite, with a value determined by $c_1$ and $c_2$. Consequently, the scalar curvature $R$ is also finite and positive during the cyclic evolution in the conformal time $\eta$, satisfying $R=4 \hat{\Lambda}$, as we have explained previously. It is precisely the fact that $\hat{\Lambda}$ is finite what allows the matching of the evolution eras. 

Let us describe this evolution in detail. First, note that because of the periodic nature of $Q(\eta)$ and $X(\eta)$, the origin of $\eta$ is irrelevant. Then, we begin the description at $Q=0$, which is marked as point $A$ in figure~\ref{fig:cyclic} and corresponds to the scale factor $a$ approaching $\infty$ (see point $A$ in figure~\ref{fig:cyclic_a}). The evolution continues as $Q(\eta)$ decreases to a minimum value at the point $B$ of figure~\ref{fig:cyclic}. This minimum means that the scale factor has reached its smallest value, $a_{\rm min}$, as we can see at point B of figure~\ref{fig:cyclic_a}. At this point, spacetime stops its collapse and begins to expand, with $a$ increasing until it approaches $\infty$ (and $Q(\eta)$ returns to zero at point $C$ of figure~\ref{fig:cyclic}). At point $C$ of figure~\ref{fig:cyclic}, we match the solutions $Q(\eta)$ and $X(\eta)$. Here, $X=0$ corresponds to $a$ going to $\infty$ (see point C of figure~\ref{fig:cyclic_a}). The evolution continues until $X(\eta)$ reaches a maximum at point $D$ of figure~\ref{fig:cyclic}, indicating that the scale factor has once again reached the minimum value $a_{\rm min}$ (see point D of figure~\ref{fig:cyclic_a}). At this point the collapse stops, and the spacetime begins a new expansion era, with $a$ growing until $a$ goes to $\infty$ (and $X(\eta)$ returns to zero at point $E$ of figure~\ref{fig:cyclic}. At this point, we then switch back to the solution $Q(\eta)$, initiating another cycle of collapse that concludes when the scale factor reaches $a_{\rm min}$, as shown in figure~\ref{fig:cyclic_a}. This periodic behavior continues indefinitely as the cosmological time $\eta$ goes on. This cyclic behavior of spacetime resembles Penrose's cycles of time~\cite{Penrose_2010}. 

\begin{figure}[hb!]
\begin{center}
\begin{tikzpicture}[xscale=0.55, yscale=0.9] 
    \definecolor{redshade}{RGB}{255, 200, 200}  
    \definecolor{blueshade}{RGB}{173, 216, 230} 
    \fill[redshade, domain=0:4*pi, variable=\x, samples=100]
        plot ({\x}, {max(0, cos(\x r))}) -- (4*pi,0) -- (0,0) -- cycle;

    \fill[blueshade, domain=0:4*pi, variable=\x, samples=100]
        plot ({\x}, {min(0, cos(\x r))}) -- (4*pi,0) -- (0,0) -- cycle;

    \draw[thick, red, domain=0:pi/2, samples=50, variable=\x] 
        plot ({\x}, {cos(\x r)});
    \draw[thick, red, domain=3*pi/2:5*pi/2, samples=50, variable=\x] 
        plot ({\x}, {cos(\x r)});
    \draw[thick, red, domain=7*pi/2:4*pi, samples=50, variable=\x] 
        plot ({\x}, {cos(\x r)});

    \draw[thick, blue, domain=pi/2:3*pi/2, samples=50, variable=\x] 
        plot ({\x}, {cos(\x r)});
    \draw[thick, blue, domain=5*pi/2:7*pi/2, samples=50, variable=\x] 
        plot ({\x}, {cos(\x r)});

    \draw[->] (-0.5,0) -- (13,0) node[below] {$\eta$};
    \draw[->] (0,-2) -- (0,2) node[left] {$X$};
    \node at (-0.5,-1.9) {$Q$};
    \node at (-0.5,-0.3) {$0$};
    
    \foreach \x in { 0.5*pi, 3*pi/2, 5*pi/2} {
        \fill (\x,0) circle (2pt); }

    \foreach \x in { 2*pi} { \fill (\x,1) circle (2pt);}
    \foreach \x in { pi} { \fill (\x,-1) circle (2pt);}

   \node at (1.1,-0.3) {$A$};
   \node at (3.1,-1.4) {$B$};
   \node at (5.1,-0.3) {$C$};
   \node at (6.3,1.4) {$D$};
   \node at (7.4,-0.3) {$E$};
   
    \draw[->] (5.5,-1.2) -- (4.5, -0.7);
    \draw[->] (7.2,-1.2) -- (8.2, -0.7);
    \node at (6.4,-1.55) {$Q=-\frac{1}{a}$};

    \draw[->] (2.2,1.2) -- (1.2, 0.7);
    \draw[->] (4,1.2) -- (5, 0.7);
    \node at (3.1,1.55) {$X=\frac{1}{a}$};
    
\end{tikzpicture}
\end{center}
  \caption{Cyclic evolution in the conformal time $\eta$ for $k=1$, by matching the solutions $Q(\eta)$ and $X(\eta)$.}
    \label{fig:cyclic}
\end{figure}
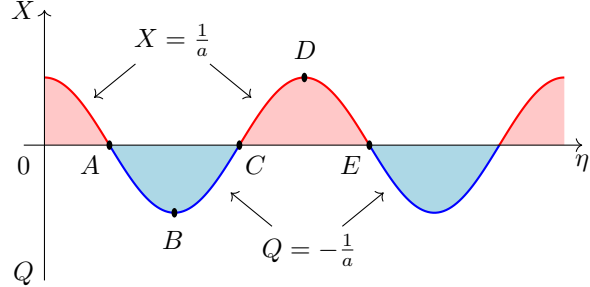

\begin{figure}[hb!]
\begin{center}
\begin{tikzpicture}[xscale=0.55, yscale=0.9] 

    \definecolor{redshade}{RGB}{255, 200, 200}  
    \definecolor{blueshade}{RGB}{173, 216, 230} 
    \fill[redshade, domain=0:pi/2-pi/11.5, variable=\x, samples=100]
        plot ({\x}, {max(0, 1/cos(\x r))}) -- (pi/2-pi/11.5,0) -- (0,0) -- cycle;

    \fill[redshade, domain=3*pi/2+pi/11.5:5*pi/2-pi/11.5, variable=\x, samples=100]
        plot ({\x}, {max(0, 1/cos(\x r))}) -- (5*pi/2-pi/11.5,0) -- (3*pi/2+pi/11.5,0) -- cycle;

     \fill[redshade, domain=7*pi/2+pi/11.5:4*pi, variable=\x, samples=100]
        plot ({\x}, {max(0, 1/cos(\x r))}) -- (4*pi,0) -- (7*pi/2+pi/11.5,0) -- cycle;
   
    \fill[redshade, domain=pi/2-pi/11.2:pi/2, variable=\x, samples=100]
        plot ({\x}, 3.7) -- (pi/2,0) -- (pi/2-pi/11.2,0) -- cycle;

    \fill[redshade, domain=3*pi/2:3*pi/2+pi/11.2, variable=\x, samples=100]
        plot ({\x}, 3.7) -- (3*pi/2+pi/11.2,0) -- (3*pi/2,0) -- cycle;
        
    \fill[redshade, domain=5*pi/2-pi/11.2:5*pi/2, variable=\x, samples=100]
        plot ({\x}, 3.7) -- (5*pi/2,0) -- (5*pi/2-pi/11.2,0) -- cycle;

    \fill[redshade, domain=7*pi/2:7*pi/2+pi/11.2, variable=\x, samples=100]
        plot ({\x}, 3.7) -- (7*pi/2+pi/11.2,0) -- (7*pi/2,0) -- cycle;

    \fill[blueshade, domain=pi/2+pi/11.5:3*pi/2-pi/11.5, variable=\x, samples=100]
        plot ({\x}, {max(0, -1/cos(\x r))}) -- (3*pi/2-pi/11.5,0) -- (pi/2+pi/11.5,0) -- cycle;
    
      \fill[blueshade, domain=5*pi/2+pi/11.5: 7*pi/2-pi/11.5, variable=\x, samples=100]
        plot ({\x}, {max(0, -1/cos(\x r))}) -- (7*pi/2-pi/11.5,0) -- (5*pi/2+pi/11.5,0) -- cycle;

    \fill[blueshade, domain=pi/2:pi/2+pi/11.1, variable=\x, samples=100]
        plot ({\x}, 3.7) -- (pi/2+pi/11.1,0) -- (pi/2,0) -- cycle;
    
    \fill[blueshade, domain=3*pi/2-pi/11.1:3*pi/2, variable=\x, samples=100]
        plot ({\x}, 3.7) -- (3*pi/2,0) -- (3*pi/2-pi/11.1,0) -- cycle;

    \fill[blueshade, domain=5*pi/2:5*pi/2+pi/11.1, variable=\x, samples=100]
        plot ({\x}, 3.7) -- (5*pi/2+pi/11.1,0) -- (5*pi/2,0) -- cycle;

    \fill[blueshade, domain=7*pi/2-pi/11.1:7*pi/2, variable=\x, samples=100]
        plot ({\x}, 3.7) -- (7*pi/2,0) -- (7*pi/2-pi/11.1,0) -- cycle;
    
    \draw[thick, red, domain=0:pi/2-pi/11.5, samples=50, variable=\x] 
        plot ({\x}, {1/cos(\x r)});
    \draw[thick, red, domain=3*pi/2+pi/11.5:5*pi/2-pi/11.5, samples=50, variable=\x] 
        plot ({\x}, {1/cos(\x r)});
    \draw[thick, red, domain=7*pi/2+pi/11.5:4*pi, samples=50, variable=\x] 
        plot ({\x}, {1/cos(\x r)});

    \draw[thick, blue, domain=pi/2+pi/11.5:3*pi/2-pi/11.5, samples=50, variable=\x] 
        plot ({\x}, {-1/cos(\x r)});
    \draw[thick, blue, domain=5*pi/2+pi/11.5:7*pi/2-pi/11.5, samples=50, variable=\x] 
        plot ({\x}, {-1/cos(\x r)});

    \draw[dashed, thick] ({pi/2}, 0) -- ({pi/2}, 3.7);
    \draw[dashed, thick] ({3*pi/2}, 0) -- ({3*pi/2}, 3.7);
    \draw[dashed, thick] ({5*pi/2}, 0) -- ({5*pi/2}, 3.7);
    \draw[dashed, thick] ({7*pi/2}, 0) -- ({7*pi/2}, 3.7);

    \draw[->] (-0.5,0) -- (13,0) node[below] {$\eta$};
    \draw[->] (0,-0.3) -- (0,3.7) node[left] {$a$};
    \node at (-0.5,-0.3) {$0$};
    
    \foreach \x in { 0.5*pi, 3*pi/2, 5*pi/2} {
        \fill (\x,0) circle (2pt); }

    \foreach \x in { 2*pi} { \fill (\x,1) circle (2pt);}
    \foreach \x in { pi} { \fill (\x,1) circle (2pt);}

   \node at (1.1,-0.3) {$A$};
   \node at (3.1,1.4) {$B$};
   \node at (5.1,-0.3) {$C$};
   \node at (6.3,1.4) {$D$};
   \node at (7.4,-0.3) {$E$};
     
\end{tikzpicture}
\end{center}
  \caption{Scale factor during the cyclic evolution in the conformal time $\eta$ for $k=1$. At points $B$ and $D$, the scale factor reaches its minimum value $a_{\rm min}$.}
    \label{fig:cyclic_a}
\end{figure}
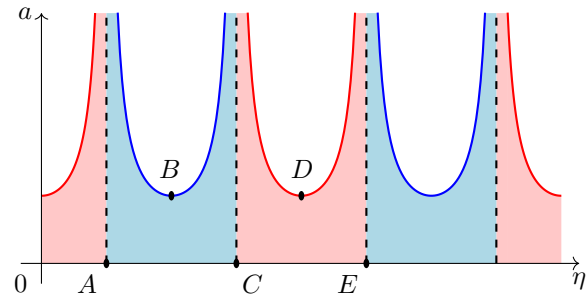

\section{Quantization}

We will now focus on performing the quantization of trace-free Einstein gravity in the FRW spacetime when there are no matter fields. Before starting, let us review the options available to carry out this task. First, we recall that, in the absence of matter fields, we have obtained both Lagrangian and Hamiltonian formulations, in both the cosmic time $t$ and the conformal time $\eta$. This allows us to choose a quantization scheme that uses either the Lagrangian description (path integral quantization) or the Hamiltonian one (canonical quantization). Due to the nature of the formulations we have found, we adopt the canonical quantization, which turns out to be simpler and more direct than its Lagrangian counterpart. Furthermore, we will proceed to quantize the Hamiltonian system~\eqref{conf_Hamil} where the evolution parameter is the conformal time $\eta$, since it avoids the ordering ambiguity present in the quantization of the Hamiltonian formulation~\eqref{CF}, and is also simpler than the Hamiltonian formulation~\eqref{CF2} that employs the cosmic time $t$ as the evolution parameter.

To quantize the system~\eqref{conf_Hamil}, we promote $Q$ and $P$ to the operators $\hat{Q}:=Q$ and $\hat{P}=-{\rm i} \hbar \frac{\partial}{\partial Q}$, and impose the commutation relation
\begin{eqnarray}
    [ \hat{Q}, \hat{P} ] = {\rm i} \hbar.
\end{eqnarray}
Then, we set up the Schr\"odinger equation
\begin{eqnarray}
    \left( -\frac{\hbar^2}{2} \frac{\partial^2}{\partial Q^2} + V(Q) \right) \Psi(Q,\eta)= {\rm i} \hbar \frac{\partial}{\partial \eta } \Psi(Q,\eta),
\end{eqnarray}
where $\Psi(Q,\eta)$ is the wave function of the system and 
\begin{eqnarray}
    V(Q) =
\left\{
\begin{array}{ll}
	\case{1}{2} k Q^2, & Q < 0, \\[6pt]
	\infty, & Q \geq 0,
\end{array}
\right.
\end{eqnarray}
with $k=\{0,\pm1\}$, as before. The Schr\"odinger equation can be solved by considering 
\begin{eqnarray}
    \Psi(Q,\eta)= e^{-{\rm i} E\eta / \hbar} \psi(Q),
\end{eqnarray}
where $E$ is a constant that allows the separation of the variables $Q$ and $\eta$, and thus $\psi(Q)$  satisfies the eigenvalue equation
\begin{eqnarray}\label{eigenvalue_eq}
     -\frac{\hbar^2}{2} \frac{d^2 \psi(Q)}{d Q^2} + V(Q)\psi(Q)  = E \psi(Q).
\end{eqnarray}
For $Q \geq 0$, where $V(Q) =\infty$, we require $\psi(Q)=0$. In the region $Q < 0$, where $V(Q) =\frac12 kQ^2$, we will consider each possible value of $k$ separately. 

\subsection{Case: $k=0$}

In this case, the system corresponds to a free particle encountering an infinite potential wall at $Q=0$. The general solution of~\eqref{eigenvalue_eq} is
\begin{eqnarray}
    \psi(Q)=C_1 \sin(\beta Q) + C_2 \cos(\beta Q),
\end{eqnarray}
where $\beta=\sqrt{2E/\hbar^2}$ with $E>0$, and $C_1$ and $C_2$ are arbitrary coefficients. However, because of the boundary condition $\psi (Q)|_{Q=0}=0$, we have $C_2=0$ and the solution simplifies to 
\begin{eqnarray}
    \psi(Q)=C_1 \sin(\beta Q).
\end{eqnarray}
Clearly, this eigenfunction is not normalizable over the infinite domain and the particle has a continuous spectrum for $E$.

\subsection{Case: $k=1$}
In this case, the system describes a quantum particle experiencing a harmonic potential for $Q<0$ and an infinite potential wall at $Q=0$ (half-harmonic oscillator). It is well known that the (normalizable) solutions   of~\eqref{eigenvalue_eq} over $Q \in (-\infty,\infty)$, which occur at the discrete values $E_m=\left(m+\frac12\right)\hbar$, correspond with the usual eigenstates of the full harmonic oscillator
\begin{eqnarray}\label{full_osc_wf}
    \psi_m(Q)=A_m e^{ -Q^2/2\hbar} H_{m} \left( \frac{Q}{\sqrt{\hbar}} \right), \qquad m=0,1,2,\dots
\end{eqnarray}
where $H_{m}(\xi)=(-1)^m e^{\xi^2} \frac{d^m}{d\xi^m}e^{-\xi^2}$ are the Hermite polynomials and $A_m$ are the normalization constants. Then, because of the boundary condition $\psi_m(Q)|_{Q=0}=0$ and the fact that $H_m(0) = 0$ if $m$ is odd, it follows that the eigenstates of the system are the odd eigenstates of the full oscillator. More precisely, the eigenfunctions normalized in the interval $Q \in (-\infty,0)$ are 
\begin{eqnarray}\label{half_osc_wf}
  \fl  \psi_n(Q)=\frac{1}{\sqrt{2^{2n}(2n+1)!\sqrt{\pi \hbar}}} e^{ -Q^2/2\hbar} H_{2n+1} \left( \frac{Q}{\sqrt{\hbar}} \right), \qquad n=0,1,2,\dots
\end{eqnarray}
which automatically satisfy the boundary condition. Furthermore, the corresponding eigenvalues are given by
\begin{eqnarray}\label{half_osc_E}
    E_n=\left( 2n + \frac{3}{2} \right) \hbar.
\end{eqnarray}

\begin{figure}
\begin{center}
		\includegraphics[width= 0.6 \columnwidth]{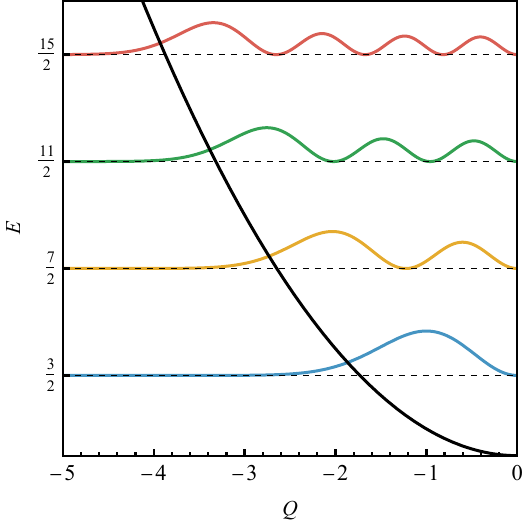}
\end{center}
\caption{Probability density $(\psi_n(Q))^2$ of the half-harmonic oscillator for $n=0,1,2,3$, together with the potential $V(Q)=Q^2/2$ as function of $Q$.}
\label{plot1}
\end{figure}

In figure~\ref{plot1} we plot the probability density $(\psi_n(Q))^2$ for the first four eigenstates, setting $\hbar=1$. Note that $(\psi_n(Q))^2$ reaches its maximum values only in the classically allowed region. On the other hand, it is also worth noting that the spectrum of the quantum observable corresponding to the cosmological constant $\hat{\Lambda}$ is
\begin{eqnarray}
\Lambda_n = \frac{6}{(l_P)^2}  \left ( 2 n + \frac{3}{2}\right ),
\end{eqnarray}
where we have used appropriate units and introduced  the Planck length $l_P$. This is one of the main results of this paper and implies that the Universe cannot reach a state where the constant of motion identified as the cosmological constant is zero. Consequently, even in the absence of matter fields, the Universe is always expanding.

Clearly, trace-free Einstein gravity and general relativity are different theories. A remarkable aspect of this departure is the fact that in trace-free Einstein gravity it is possible to construct a quantum observable for the classical constant motion $\hat{\Lambda}$ identified as the cosmological constant. In general relativity, the cosmological constant is a parameter present in the classical equations of motion or in the action principle that yields them. Therefore, it makes no sense to construct a quantum observable for the cosmological constant in general relativity.

{\it Heisenberg's uncertainty principle}. Since a quantum system cannot be completely localized, it is to be expected that there exists a limit to simultaneous measurements of position and momentum. This restriction is expressed by Heisenberg's uncertainty relation
\begin{eqnarray}\label{Heisenberg_uncert}
    \Delta Q \, \Delta P \geq \frac{\hbar}{2},
\end{eqnarray}
where $(\Delta Q)^2=\langle \hat{Q}^2\rangle_n-\langle \hat{Q}\rangle^2_n$ is the variance of the operator $\hat{Q}$ in the quantum state $\psi_n$, and $(\Delta P)^2$ is defined analogously. Using~\eqref{half_osc_wf}, the left-hand side of~\eqref{Heisenberg_uncert} is
\begin{eqnarray}\label{Heisenberg2}
    \Delta Q  \, \Delta P=\sqrt{4n+3\left(4n+3-\frac{32 (\Gamma(n+3/2))^2}{\pi^2 (n!)^2} \right) } \frac{\hbar}{2}.
\end{eqnarray}
In particular, for $n=0,1,2$ this expression yields $\Delta Q  \, \Delta P \approx 0.58\hbar,1.49 \hbar, 2.37 \hbar $, respectively, which is consistent with~\eqref{Heisenberg_uncert}.

 It is important to understand the implications of equations~\eqref{Heisenberg_uncert} and~\eqref{Heisenberg2}. For this purpose, we recall that in the current context, the quantum operator $\hat{P}$ is associated with the classical momentum $P$, which is precisely the expansion rate $H$ [see equation~\eqref{exp_rate}], and that $\hat{Q}$ is the quantum operator corresponding to the classical variable $Q$, which is related to the scale factor $a$ as defined by~\eqref{conf_defQ}. Therefore, the relation~\eqref{Heisenberg_uncert} implies an intrinsic limitation in simultaneous knowledge of the expansion rate and the scale factor: both quantities cannot be measured simultaneously with arbitrary precision. Note that as a consequence of this fact, the minimum value of $E_n$ (and therefore of $\Lambda_n$) is not zero. Moreover, since the scale factor is directly related to the cosmological redshift $z$, this limitation extends to the measurement accuracy of $z$ as well.

\subsection{Case: $k=-1$} 

In this case, the system corresponds to a quantum particle subject to an inverted parabolic potential in the region $Q<0$, which acts as the repulsive potential, and confined by an infinite potential wall at $Q=0$ (half-repulsive oscillator). It can be shown that the general solution of~\eqref{eigenvalue_eq} over $Q \in (-\infty,\infty)$ is (see the appendix)
\begin{eqnarray}\label{gral_sol_repul}
   \fl \psi(q) = e^{-{\rm i}Q^2/2\hbar} \bigg[ C_1 \, {}_1F_1 \left( \frac{{\rm i} E}{2\hbar} + \frac14; \frac{1}{2}; \frac{{\rm i Q^2}}{\hbar} \right) + C_2 \, Q \, {}_1F_1 \left( \frac{{\rm i} E}{2\hbar}+\frac{3}{4}; \frac{3}{2}; \frac{{\rm i Q^2}}{\hbar} \right) \bigg],
\end{eqnarray}
where ${}_1F_1(a;b;z)$ is the confluent hypergeometric function of the first kind
\begin{eqnarray}
    {}_1F_1(a;b;z)=\frac{\Gamma(b)}{\Gamma(a)} \sum_{l=0}^{\infty} \frac{\Gamma(a+l)}{\Gamma(b+l)} \frac{z^l}{l!},
\end{eqnarray}
with $\Gamma$ being the standard gamma function; $C_1$ and $C_2$ are arbitrary coefficients. Taking into account the boundary condition $\psi (Q)|_{Q=0}=0$ and the fact that $_1F_1(a;b;0)=1$, we have $C_1=0$ and then the solution reduces to
\begin{eqnarray}
    \psi(q) = C_2 \, Q \, e^{-{\rm i}Q^2/2\hbar} {}_1F_1 \left( \frac{{\rm i} E}{2\hbar}+\frac{3}{4}; \frac{3}{2}; \frac{{\rm i Q^2}}{\hbar} \right).
\end{eqnarray}
In this expression, $E$ can take any value, resulting in a continuous spectrum. It is worth noticing that $\psi(q)$ is real if $C_2$ is real. Additionally, it can be verified that $\psi(Q)\rightarrow 0$ as $Q\rightarrow-\infty$. Nevertheless, this solution exhibits oscillations that become progressively faster and cannot be normalized over the entire region $Q<0$.

\begin{figure}
\begin{center}
		\includegraphics[width= 0.6 \columnwidth]{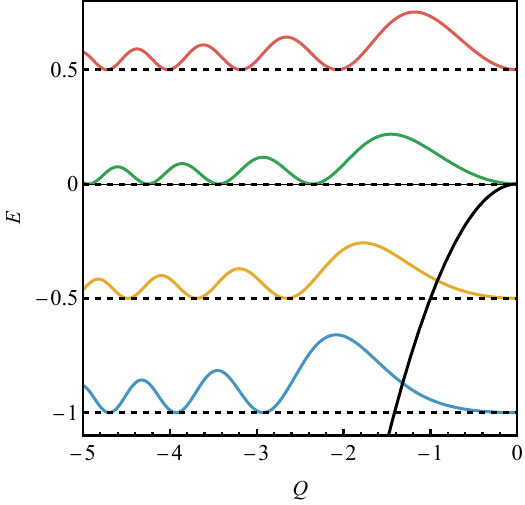}
\end{center}
\caption{Probability density $(\psi(Q))^2$ of the half-repulsive oscillator for $E=-1,-0.5,0,0.5$, together with the potential $V(Q)=-Q^2/2$ as function of $Q$.}
\label{fig2}
\end{figure}

In figure~\ref{fig2} we plot the probability density $(\psi(Q))^2$ for some values of $E$, assuming $\hbar=1$ and $C_2$ to be real. Note that the maxima of the probability density are confined to region before the particle reaches the repulsive potential. 

\section{Conclusions}

We have investigated, in the context of trace-free Einstein gravity and using the FRW metric~\cite{Ellis_2011,Ellis_2014}, the consequences of the fact that the cosmological constant is a constant of motion $\hat{\Lambda}$, rather than a fixed universal (yet unknown) parameter $\Lambda$ as in general relativity. Our results force us to rethink the role and meaning of the cosmological constant. Although this aspect is often overlooked in trace-free Einstein gravity~\cite{Einstein_1919,Einstein_1952,Einstein_1927} (and similarly in unimodular gravity~\cite{Anderson_1971,Weinberg_1989}), the findings presented in this paper indicate that it deserves more careful exploration.

In the absence of matter fields, and using the conformal time as time variable (evolution parameter), trace-free Einstein equations reduce to a single equation that has the form of a free particle (flat Universe), a harmonic oscillator (closed Universe), and a repulsive oscillator (open Universe) in the real half-line. In the classical theory, we have found that the scalar curvature, which is on-shell $4\hat{\Lambda}$, remains finite along the evolution and thus there are no singularities. This fact should not be confused with what happens in general relativity with cosmological constant $\Lambda$ (de Sitter) because in our case (trace-free Einstein gravity) the observable identified as the cosmological constant, $\hat{\Lambda}$, arises dynamically as a consequence of the equations of motion and is not given a priori as a fixed parameter. Furthermore, the case $k=0$ analyzed here should also not be confused with the case studied in~\cite{Perez_2023}, where the time variable employed is not the conformal time and the theory is unimodular gravity. 

We have also shown that the constant of motion $\hat{\Lambda}$ plays a fundamental role in the Hamiltonian formulations given, as in each case the Hamiltonian is proportional to it. This implies, in the quantum theory, that it is possible to calculate the spectrum of the observable associated with the cosmological constant, which turns out to be discrete and positive for the closed Universe and real for the flat and open universes. This finding means that the Universe cannot evolve into a state where this constant is zero; as a result, even in the absence of matter, the Universe continues to expand. The discreteness of the observable corresponding to the cosmological constant is a major departure from general relativity, where it makes no sense to calculate the spectrum of the cosmological constant because there it is a parameter, not a constant of motion of the theory. It is important to emphasize that our analysis has been carried out in vacuum (i.e., without matter fields). Therefore, it would be desirable to obtain the spectrum of the cosmological constant in the presence of matter fields as well. Further research in this direction is required. It is worth noting that, to the best of our knowledge, the only other theory of gravity allowing for the absence of singularities is loop quantum gravity~\cite{Bojowald_PRL}.

Moreover, in the quantum theory, the observables corresponding to the phase space variables must satisfy Heisenberg’s uncertainty relation. Consequently, simultaneous measurements of the momentum variable, which is the Hubble expansion (Hubble parameter), and the configuration variable, which is the negative or the positive of the inverse of the scale factor, are inherently constrained by Heisenberg's uncertainty principle. Such a fundamental limitation arising from Heisenberg's uncertainty principle should, in principle, be observationally measured if the theory is physically right.

It is important to emphasize that the quantization of the FRW cosmology of trace-free Einstein gravity reported in this paper is not diffeomorphism-invariant but represents the quantum theory from the perspective of an observer using conformal time as the evolution parameter. It would be worthwhile to compare the current result with a diffeomorphism-invariant quantization of the theory, which does not exist as far as we know. In particular, the confrontation with a diffeomorphism-invariant quantum theory that emerges from the diffeomorphism-invariant actions~\cite{MontGonz_2023,MontGonz_2024,Montesinos_2025} is something mandatory. Finally, the quantization of the trace-free Einstein cosmology in the FRW scenario using the various canonical and noncanonical Hamiltonian formulations reported in this paper is also worthwhile.

\ack 

We thank Alejandro Perez, Ulises Nucamendi, and Mariano Celada for fruitful discussions on the subject. Diego Gonzalez acknowledges the financial support of Instituto Politécnico
Nacional, Grant No. SIP-20253696, and the postdoctoral fellowship from Consejo Nacional de Humanidades, Ciencia y Tecnolog\'{\i}a (CONAHCyT), M\'exico.


\appendix 

\section{Eigenfunctions of the repulsive oscillator}
While the derivation of the harmonic oscillator eigenfunctions is well-known and widely available, the derivation of the repulsive oscillator eigenfunctions is not. Here, we  fill out this gap.  For a different approach based on Whittaker functions, see~\cite{Wolf_book}. 

By performing some change of variables, the equation~\eqref{eigenvalue_eq} with $V(Q) =-\frac12 Q^2$, which is rewritten as 
\begin{eqnarray}\label{appx_eigenvalue_eq}
 \frac{d^2 \psi(Q)}{d Q^2} + \frac{1}{\hbar^2} \left( Q^2 + 2E \right) \psi(Q)  = 0,
\end{eqnarray}
can be cast into the confluent hypergeometric equation or Kummer's equation. This is done as follows. First, by introducing either the variable $Z=(1+{\rm i}) Q/\sqrt{\hbar}$ or $Z=-(1+{\rm i}) Q/\sqrt{\hbar}$, the previous equation becomes
\begin{eqnarray}
    \frac{d^2 \psi(Z)}{d Z^2} - \left( \frac{{\rm i} E}{\hbar} + \frac{Z^2}{4} \right) \psi(Z)  = 0,
\end{eqnarray}
which has the form of the Weber equation. Then, with the substitution $\psi(Z)=e^{-Z^2/4} u(Z)$, where $ u(Z)$ is a function of $Z$ to be determined, we get
\begin{eqnarray}
    \frac{d^2 u(Z)}{d Z^2} - Z \frac{d u(Z)}{d Z} - \left(\frac{{\rm i} E}{\hbar}+\frac12\right) u(Z) = 0.
\end{eqnarray}
The further change of variable $\tau=Z^2/2$ converts this equation into
\begin{eqnarray}
    \tau \frac{d^2 u(\tau)}{d \tau^2} + \left( \frac12 - \tau\right) \frac{d u(\tau)}{d \tau} - \left(\frac{{\rm i} E}{2\hbar}+\frac14\right) u(\tau) = 0.
\end{eqnarray}
This is the form of the confluent hypergeometric equation or Kummer's equation, whose general solution is~\cite{Willatzen_book}
\begin{eqnarray}
    u(\tau)=&K_1 \, {}_1F_1\left(\frac{{\rm i} E}{2\hbar}+\frac14;\frac12;\tau\right) +K_2 \, \tau^{1/2} {}_1F_1\left(\frac{{\rm i} E}{2\hbar}+\frac{3}{4};\frac{3}{2};\tau\right),
\end{eqnarray}
where $K_1$ and $K_2$ are arbitrary constant coefficients. Therefore, the corresponding solution of~\eqref{appx_eigenvalue_eq} is $\psi(Q)=e^{-{\rm i}Q^2/2\hbar} u(\tau)$ with $\tau={\rm i} Q^2/ \hbar$, which is precisely~\eqref{gral_sol_repul} after a redefinition of constants (see also~\cite{Yuce_2021}). Note that $E$ need not be an integer.


\section*{References}
\bibliography{references}

\end{document}